# Genetic drift suppresses bacterial conjugation in spatially structured populations

**Peter D. Freese,**[†] **Kirill S. Korolev,**[†‡§¶] **José I. Jiménez,**[† ‖] **and Irene A. Chen**[†‖‖*]

[†]FAS Center for Systems Biology, Harvard University, Cambridge, Massachusetts; [‡]Department of Physics, Harvard University, Cambridge, Massachusetts; [§]Department of Physics, Massachusetts Institute of Technology, Cambridge, Massachusetts; [¶]Department of Physics and Program in Bioinformatics, Boston University, Boston, Massachusetts; [‖]Faculty of Health and Medical Sciences (University of Surrey, UK) and [‖‖]Department of Chemistry and Biochemistry, Program in Biomolecular Sciences, University of California, Santa Barbara, California

*Correspondence: chen@chem.ucsb.edu

**Contents**

**ABSTRACT**

Conjugation is the primary mechanism of horizontal gene transfer that spreads antibiotic resistance among bacteria. Although conjugation normally occurs in surface-associated growth (e.g., biofilms), it has been traditionally studied in well-mixed liquid cultures lacking spatial structure, which is known to affect many evolutionary and ecological processes. Here we visualize spatial patterns of gene transfer mediated by F plasmid conjugation in a colony of *Escherichia coli* growing on solid agar, and we develop a quantitative understanding by spatial extension of traditional mass-action models. We found that spatial structure suppresses conjugation in surface-associated growth because strong genetic drift leads to spatial isolation of donor and recipient cells, restricting conjugation to rare boundaries between donor and recipient strains. These results suggest that ecological strategies, such as enforcement of spatial structure and enhancement of genetic drift, could complement molecular strategies in slowing the spread of antibiotic resistance genes.

# INTRODUCTION

Antibiotics are one of the most important medical interventions of the last century. Yet the extensive use of antibiotics selects for resistance among pathogenic bacteria, which already limits treatment of some major types of infection (1). The increase in resistance is primarily driven by the spread of resistance genes already present in natural communities. A major mechanism for horizontal gene transfer is bacterial conjugation (2), which has spread resistance to β-lactams and aminoglycosides to clinically significant organisms (3). The important role of conjugation in the spread of antibiotic resistance, and in microbial evolution in general, motivates both fundamental study of conjugation and strategies to inhibit it.

Conjugation requires physical contact between a donor and recipient cell. The donor cell carries a conjugative plasmid, which contains genes necessary for conjugation and possibly other genes (e.g., encoding antibiotic resistance). A competent donor cell expresses a pilus, which binds to the recipient cell and mediates plasmid DNA transfer. For example, in the well-studied F factor system, an $F^+$ donor cell transfers the plasmid to the recipient, initially $F^-$, cell, thus creating a new $F^+$ transconjugant cell. Conjugation and maintenance of the plasmid slightly reduces organismal fitness, and a large fitness cost is paid in the presence of certain phages (e.g., the filamentous phages including M13, fd, and f1), which attach specifically to the conjugative pilus. Indeed, addition of M13 or its attachment protein, g3p, reduces the rate of conjugation from $F^+$ cells and could be an interesting strategy to suppress undesired horizontal gene transfer (4, 5).

Many clinically and environmentally important habitats are spatially structured because bacteria live in surface-associated colonies and biofilms, where motility is limited (6, 7). Although spatial structure is known to play an important role in evolutionary dynamics, its effect on conjugation dynamics has been largely unexplored. Indeed, modeling and experimental studies of conjugation have previously focused on simple well-mixed liquid systems like batch cultures and chemostats, which can be described by mass-reaction equations. These traditional approaches neglect important aspects of natural populations that result from spatial structure. More recently, experimental and theoretical efforts have been directed at studying conjugation in spatially structured environments. Some studies show that conjugation can be quite prevalent in a biofilm (8, 9), but others suggest that spread of conjugative plasmids in biofilms and on agar surfaces is quite limited (10, 11). When plasmid-bearing cells provide a 'public good' (e.g., by detoxifying $Hg^{2+}$ from their surroundings), the relative frequency of plasmid-bearing and plasmid-free cells also influences the fitness advantage of the plasmid (12). Interpretation of results in spatially structured environments has also been hampered by the difficulty of distinguishing among donor, recipient, and transconjugant cells by microscopy. In general, previous methods have not been able to resolve two of the three cell types in situ (13). Thus, these experimental results, which are affected by multiple factors (e.g., cell densities, plasmid characteristics, and the spatial scale of structuring), point toward a need for improved experimental systems as well as a quantitative theoretical framework to advance our fundamental understanding of conjugation.

Models developed to describe homogeneous environments do not properly capture dynamics on heterogeneous environments (14-16). Early models of conjugation in spatially structured environments included unrealistic assumptions or did not allow measurement of transfer events per donor-recipient encounter, which is necessary for comparison of conjugation rates across different species and situations (13, 17, 18). A more recent spatial model of conjugation (19) used cellular automata to simulate individual cells in a lattice and captured features of experimental conjugation. Although this analysis considered the possibility of local plasmid extinction, it was unable to completely determine genetic history during colony expansion experiments, as donors and transconjugants could not be discriminated in situ. This history is important because in spatially structured populations, only a small number of nearby cells compete with each other, leading to substantial demographic stochasticity (genetic drift) on short spatial scales (20). As a result, some genotypes become extinct locally, leading to a macroscopic pattern of isogenic domains (sectors) in a growing bacterial colony. The number of sectors tends to decrease over time because sectors irreversibly disappear due to genetic drift when sector boundaries cross. We hypothesized that this spatial demixing of genotypes could profoundly affect bacterial conjugation because plasmid transfer requires spatial proximity of the donor and recipient cells.

In this study, we explored the dynamics of bacterial conjugation in colonies grown on an agar surface by combining spatially resolved measurements and simulations. We visualized the spatial distribution of donor cells ($F^+$ encoding tetracycline resistance ($Tc^r$), with cyan fluorescent protein expressed from a non-conjugative plasmid) and recipient cells (initially $F^-$, tetracycline sensitive ($Tc^s$), with yellow fluorescent protein expressed from a non-conjugative plasmid). Since the $Tc^r$ phenotype is carried on the F plasmid, transconjugants are $Tc^r$ yellow fluorescent cells. Populations of transconjugants could be visually distinguished from $F^-$ cells by a decrease in fluorescence intensity caused by partial repression of the fluorescent protein, as well as by tetracycline resistance. Contrary to the general belief that biofilms facilitate conjugation, we found that conjugation is substantially suppressed in surface growth compared to liquid culture, consistent with simulations of conjugation dynamics. Thus, spatial structure itself could be an important factor in slowing down the spread of antibiotic resistance. In addition, previous studies in liquid culture showed that exogenous addition of M13 phage particles or the soluble portion of the M13 minor coat protein g3p (g3p-N) results in nearly complete inhibition of conjugation (4, 5). We found a similar inhibitory effect of g3p-N in surface-associated bacterial colonies. The work presented here adds to prior experimental and theoretical studies of conjugation on spatially structured environments by quantifying genetic drift, which accounts for the limited penetration of the F plasmid into the spatially structured environment, and by analyzing conjugation rates in the presence of an inhibitory agent. The results suggest that molecular anti-conjugation strategies could generalize to natural spatially structured populations.

**MATERIALS AND METHODS**

**Bacterial strains and culture**

*E. coli* TOP10F' and TOP10 (Invitrogen, Carlsbad, CA; mcrA Δ(mrr-hsdRMS-mcrBC) Φ80lacZΔM15 ΔlacX74 recA1 araD 139 (ara leu) 7697 galU galK rpsL (StrR) endA1 nupG) were used as donor ($F^+_c$) or recipient ($F^-$), respectively of the F' plasmid [$lacI^q$, Tn10($tet^R$)]. Plasmids pTrc99A-eYFP ($amp^R$) and pTrc99A-eCFP ($amp^R$) encoding, respectively, eYFP (Q95M; yellow) and eCFP (A206K; cyan) were courtesy of Howard Berg. The fluorescent markers were expressed from $amp^R$ selectable plasmids under an IPTG-inducible promoter.

Standard protocols were used for common bacterial and phage-related procedures (21). All strains were grown in Luria-Bertani (LB) medium on a regular basis. When needed, media were supplemented with ampicillin or carbenicillin (100 μg/mL), tetracycline (12 μg/mL), and isopropyl-β-D-1-thiogalactopyranoside (IPTG) (1mM). Agar plates were made with 20 mL LB supplemented with carbenicillin and IPTG. Separate 3 mL overnight cultures of TOP10F' (transformed by pTrc99A-eCFP) and Top10 (transformed by pTrc99A-eYFP) were inoculated in 3 mL of LB with ampicillin, ITPG, and tetracycline if appropriate. Overnight cultures were inoculated from a colony grown on medium with the appropriate antibiotic(s) to select for the desired plasmids, and grown at 37°C with shaking at 250 rpm overnight to saturation ($OD_{600}$ ~ 3-4 determined by an Ultrospec cell density meter (GE Healthcare, Pittsburgh, PA)). $F^+$ cells grown in medium supplemented with antibiotics were centrifuged for 5 minutes at 5000 rpm, the supernatant was discarded, and the pellet was resuspended in fresh LB medium lacking antibiotics. Cultures were diluted for density measurement with appropriate medium to bring the OD$_{600}$ within the linear range of the cell density meter (i.e., <1 ODU). Strains were then mixed to the desired ratio as measured by optical density to create the inoculant. A small volume of inoculant (1-20 μL) was pipetted onto the center of an LB-agar plate containing carbenicillin and IPTG. The plates were then incubated for the desired length of time at 37°C in a bin containing wet paper towels to maintain high humidity.

**Detection of transconjugants**

After the desired growth period, transconjugants were detected by applying a tetracycline-soaked ring around the bacterial colony. The center of a 2.5 cm diameter filter paper disk (VWR, Bridgeport, NJ) was removed to create a thin annulus with inner diameter 1.9 cm. Tetracycline stock at 12 mg/mL was diluted to 2 mg/mL with 50% ethanol. 30 μL of the tetracycline mixture was applied uniformly onto an autoclaved filter paper annulus, which was then placed around a growing bacterial colony with sterilized forceps. Since the number of sector boundaries is not very large, the number and size of transconjugant sectors is expected to vary from colony to colony even though there are millions of cells growing on a Petri dish. We indeed observed much higher variability in spatial compared to liquid assays of conjugation and performed measurements on 10-100 colonies in each experiment to obtain reliable estimates of the averages.

**Application of g3p-N**

The g3p-N protein was prepared as described in (5). 4.9 μL of g3p-N stock solution (41

μM) was mixed with 45 μL of phosphate buffered saline (PBS) per plate and a 50 μL aliquot was spread on each plate with glass beads for 1-2 minutes until dry. Assuming uniform diffusion throughout the 20 mL agar plate, the expected [g3p-N] is 10 nM, a concentration which gives 80% conjugation inhibition in liquid culture (5).

**Microscopy and image processing**

Fluorescent images were obtained with a Zeiss Lumar V.12 fluorescence stereoscope (Oberkochen, Germany) and a Typhoon TRIO variable-mode imager (GE Healthcare, Pittsburgh, PA). Scanned plates were imaged from the bottom using cyan laser excitation and detection at 488 nm with 50 μm resolution. The initial radii of the colonies were measured within an hour of inoculation by fitting of a circle using the stereoscope's software; colonies that were not circular were discarded. The number of sectors in each colony was counted manually. MATLAB R2010 (The MathWorks, Natick, MA) was used to extract the radii and sector boundaries of the colonies using the built-in "edge" function.

**Modeling and simulations of conjugation in bacterial colonies**

We formulated a minimal model of surface-associated populations. Following the stepping-stone model of Kimura and Weiss (22), spatially structured populations are often modeled as an array of well-mixed populations (demes) that exchange migrants. Previous work demonstrated that genetic demixing in growing bacterial colonies can be described by a one-dimensional stepping-stone model because growth occurs only close to the nutrient-rich circumference of the colony (23, 24). Here, we formulated a model that, in addition to competition, genetic drift, and migration, incorporates horizontal gene transfer between cells. Since previous work showed that the qualitative behavior of linearly and radially expanding populations is quite similar, and both types of expansions lead to sector formation (23), we, for simplicity, neglected the fact that the circumference of the colony and therefore the total population size were changing during the experiment.

In our study, simulated populations were composed of a linear set of $L_{sim}$ demes containing $N$ cells of three possible types: F$^-$, F$^+_c$, and transconjugants with respective proportions $f^-$, $f^+_c$, and $f^t$. Each deme was treated as a well-mixed population. To account for daughter cells being displaced slightly from parent cells during colony growth, cells could migrate to one of their two nearest neighbor demes with probability $m$ per generation. Reproduction and conjugation were modeled through a series of time steps at which only two cells were updated, always preserving the total population size. In reproduction events, one individual died (or fell behind the expanding front in the context of our experiments), allowing another individual to reproduce and thus keeping the population size constant. A series of $N$ time steps corresponded to one generation because every individual was replaced once on average. Possible composition-changing events are given below with their corresponding probabilities $P$, which depend on the fitness cost $s$ of the F plasmid, conjugation rate $r$, and the local proportions of the cell types F$^-$ ($f^-$), original cyan F$^+_c$ ($f^+_c$), and transconjugant ($f^t$). These probabilities (Eqs. 1-6 below) were formulated assuming that conjugation and competition occur at a fixed probability per cell-cell interaction within each deme. Conjugation events decrease the F$^-$ population and increase the transconjugant population, while competition decreases the F$^+$
10

populations (donor strain and transconjugants) and increases the F⁻ population because the F plasmid imposes a fitness cost. For example in Eq. 1 below, the probability that the F⁻ population increases by one cell and the F⁺$_c$ population decreases by one cell is proportional to the probability of F⁻ & F⁺$_c$ interaction given by the product of their proportions $(f^- * f^+_c)$ and the sum of three terms describing genetic drift (factor of 1), competition $\left(\frac{s}{2}\right)$, and conjugation $\left(-\frac{r}{2}\right)$. As expected, this probability increases as the fitness cost of the F plasmid increases, and decreases as the conjugation rate increases.

$$P\left(f^- + \tfrac{1}{N}, f^+_c - \tfrac{1}{N}, f^t\right) = f^- f^+_c \left(1 + \tfrac{s}{2} - \tfrac{r}{2}\right) \tag{1}$$

$$P\left(f^- - \tfrac{1}{N}, f^+_c + \tfrac{1}{N}, f^t\right) = f^- f^+_c \left(1 - \tfrac{s}{2} - \tfrac{r}{2}\right) \tag{2}$$

$$P\left(f^- + \tfrac{1}{N}, f^+_c, f^t - \tfrac{1}{N}\right) = f^- f^t \left(1 + \tfrac{s}{2} - \tfrac{r}{2}\right) \tag{3}$$

$$P\left(f^- - \tfrac{1}{N}, f^+_c, f^t + \tfrac{1}{N}\right) = f^- f^+_c r + f^- f^t \left(1 - \tfrac{s}{2} + \tfrac{r}{2}\right) \tag{4}$$

$$P\left(f^-, f^+_c + \tfrac{1}{N}, f^t - \tfrac{1}{N}\right) = f^t f^+_c \tag{5}$$

$$P\left(f^-, f^+_c - \tfrac{1}{N}, f^t + \tfrac{1}{N}\right) = f^t f^+_c \tag{6}$$

In the limit of infinite population size, when fluctuations can be neglected, one can obtain a simple description of the dynamics in terms of ordinary differential equations. The key idea is to compute the average change in the relative proportions of the different cell types using Eqs. 1-6 and then treat $f^-$, $f^+_c$, and $f^t$ as deterministic variables. For example, the change of $f^-$ per one time step ($\frac{1}{N}$ of generation time) is given by:

$$E\left[\frac{1}{N}\left(P\left(f^- +\frac{1}{N}, f^+_c -\frac{1}{N}, f^t\right) - P\left(f^- -\frac{1}{N}, f^+_c +\frac{1}{N}, f^t\right) + P\left(f^- +\frac{1}{N}, f^+_c, f^t -\frac{1}{N}\right) - P\left(f^- -\frac{1}{N}, f^+_c, f^t +\frac{1}{N}\right)\right)\right]$$

$$= \frac{1}{N}\left(f^- f^+_c\left(1+\frac{s}{2}-\frac{r}{2}\right) - f^- f^+_c\left(1-\frac{s}{2}-\frac{r}{2}\right) + f^- f^t\left(1+\frac{s}{2}-\frac{r}{2}\right) - \left(f^- f^+_c r + f^- f^t\left(1-\frac{s}{2}+\frac{r}{2}\right)\right)\right) = \frac{1}{N}(s-r)f^-(t)[f^+_c(t) + f^t(t)],$$

which is the combined effect of the four possible transitions that change the number of F⁻ cells. After dividing by the time interval $\frac{1}{N}$, and repeating the same calculation for $f^+_c$ and $f^t$, we obtain the following set of differential equations:

$$\frac{d}{dt} f^-(t) = (s-r)f^-(t)[f^+_c(t) + f^t(t)] \quad (7)$$

$$\frac{d}{dt} f^+_c(t) = -sf^-(t) f^+_c(t) \quad (8)$$

$$\frac{d}{dt} f^t(t) = (r-s)f^-(t)f^t(t) + rf^-(t) f^+_c(t) \quad (9)$$

The terms in Eq. 7 allow straightforward interpretation: F⁻ cells increase due to selection at rate *s* and decrease due to conjugation at rate *r*, both of which occur proportional to the frequency at which F⁻ and F⁺ cells come together. Equations 8 and 9 allow analogous interpretation in terms of appropriate events that change the number of F⁺$_c$ and transconjugant cells.

Multiple choices of transition probabilities lead to the same behavior in the limit of infinite population size. However, the choice does not affect the dynamics provided that *s<<1* and *r<<1*, as is reasonable here, and that the dynamics are equivalent to the Moran model (25) of three neutral species for $s = r = 0$. The reason is that the full dynamics of the cells can be described by stochastic differential equations with the deterministic terms given by Eqs. 7-9 while dependence of the stochastic terms on $s$ and $r$ can be neglected.

After the $N$ time steps of reproduction and conjugation events, migration was implemented such that demes were chosen for migration in random order. When a deme was chosen, each of the $N$ individuals was sequentially selected and migrated to the right deme with probability $m/2$ and to the left deme with probability $m/2$. If the individual was chosen to migrate, a random individual from the destination migrated back to the origin so that the population size in each deme was always conserved. To avoid edge artifacts and to mimic the actual experiments, we imposed periodic boundary conditions so that a cell could migrate from the last deme to the first deme and vice versa. Further details of the model are provided in the Methods and Supporting Data: Simulation Details.

**Parameterizing the model**

In order to use the model for quantitative predictions, we parameterized the model using experimental data. We used the model to estimate the effective conjugation rate in our experimental populations by finding the model parameters that lead to the same spatial distribution of $F^-$, $F^+_c$, and transconjugant cells as observed in the experiments. The process of parameterizing the model is quite straightforward provided the following three issues are taken into account. First, spatial patterns are stochastic in both simulations and experiments, so the model should fit average properties of these patterns rather than patterns themselves. Second, not all parameters in the simulation can be determined uniquely because the choice of spatial and temporal scales in the simulations depends on the level of desired precision or coarse-graining that can be freely adjusted. Third, our simulations have a constant length, while the circumference of the colonies increases with time, so the process of comparing the patterns has to take these differences in the geometry into account. We now briefly outline the parameter fitting procedure (also see Table 1); for the complete description see the Supporting Methods: Modeling Details.

The spatial patterns that result from genetic drift and competition have been previously investigated in (23, 24, 26), where the authors showed that population dynamics without conjugation that study here can be described in terms of three quantities, the effective diffusion constant $D_s \sim m * (deme\ size)/(generation\ time)$, effective strength of genetic drift $D_g \sim (deme\ size)/(N * generation\ time)$, and outward bending of more fit sectors $v_\perp \sim s * (deme\ size)/(generation\ time)$. They have computed various statistics based on the spatial patterns of genetic demixing and competition that can be used to estimate these three parameters. Here we omit the derivations of their published results and only provide the mathematical expressions used in the analysis.

**Quantifying migration, genetic drift, and the fitness cost of the F plasmid**

Procedures for quantifying migration, genetic drift, and the fitness cost of the F plasmid are given in Supporting Material: Modeling Details.

**Connecting experiments with simulations**

The experiments have definite physical measures of time and space, but these are arbitrarily scaled in simulations. For computational efficiency, we used this freedom in choosing spatial and temporal scales to select certain values of m and N (N = 100, mN = 5) and then determined the corresponding spatial and temporal scales by matching experimental and simulation data. As we show in the Supporting Data: Simulation Details, this choice does not affect our estimate of the conjugation rate, which we further verified by repeating model parameterization for different values of m and N (N = 30, mN = 1 and N = 30, mN = 10). To match experimental and simulation data, we defined four dimensionless quantities (invariants, *Inv*) derived from the six experimental parameters $D_g$, $D_s$, $v_\perp$, $<f^t>$ (average fraction of transconjugants), $T_{exp}$ (total time), and $L_{exp}$ (population front length):

$$Inv_1 = \frac{D_s}{D_g^2 T_{exp}}, \tag{10}$$

$$Inv_2 = \frac{D_s}{D_g L_{exp}}, \tag{11}$$

$$Inv_3 = \frac{v_\perp T_{exp}}{L_{exp}}, \tag{12}$$

$$Inv_4 = <f^t>. \tag{13}$$

To establish a match, the values of these experimental invariants and their simulation counterparts must be equal. In particular, the first two invariants were used to find the number of simulation generations and demes ($T_{sim}$ and $L_{sim}$, respectively). The third invariant was used to estimate the fitness cost of the plasmid, and the fourth invariant to estimate the conjugation rate.

## RESULTS

**Visualizing conjugation during colony expansion**

To visualize conjugation, we began experiments with $F^+$ donor cells expressing eCFP (enhanced cyan fluorescent protein) and $F^-$ recipient cells expressing eYFP (enhanced yellow fluorescent protein). The two strains were grown to saturation overnight, mixed to the desired proportion (generally 1:1 $F^+$: $F^-$) by optical density, inoculated onto agar plates in drops of 1-20 µL, and incubated at 37°C. The schematic of the experiment is shown in Fig. 1*A* and expansion rates in Fig. S1 in the Supporting Material. The spatial distribution of $F^+$ donor cells and initially $F^-$ cells was visualized by fluorescence microscopy. However, both transconjugant cells and $F^-$ cells express eYFP, so we applied a ring of filter paper soaked in tetracycline to identify the $Tc^r$ transconjugants. Since only $F^+$ cells are able to grow in the presence of tetracycline ($tet^R$ being carried on the F plasmid), transconjugant sectors appeared as yellow fluorescent sectors that continued to grow after the application of tetracycline (Fig. 1*B*). In the following, we refer to cyan fluorescent $F^+$ cells as $F^+_c$, yellow fluorescent $F^-$ cells as $F^-$, and yellow fluorescent $F^+$ cells as transconjugants.

We observed that the application of tetracycline caused a decrease in the fluorescence intensity of transconjugant cells, as illustrated by the difference between transconjugant and $F^-$ cells (Fig. 1*B*). This phenomenon is because the $lacI^q$ carried on the F plasmid partially represses expression of the fluorescent protein, which is under an IPTG-inducible promoter. This intensity effect enables visualization of the boundaries of the transconjugant sector so the transconjugant can be traced back to its origin, presumably close to the conjugation event.

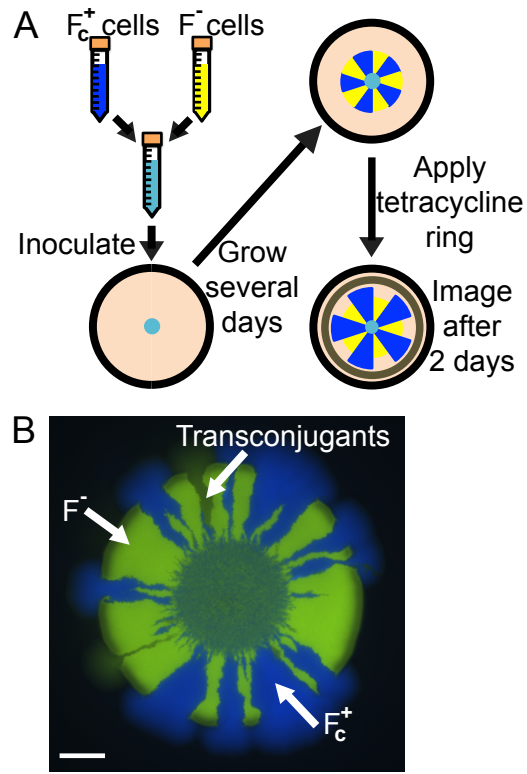

FIGURE 1 Experimental setup. (*A*) $F^-$ and $F^+_c$ liquid cultures were grown to saturation overnight, mixed to the desired ratio (most often 1:1) as measured by optical density, and 1-20 μL was pipetted onto an agar plate. After 4-7 days of growth, a ring of tetracycline was applied around the colony, which diffused through the agar and allowed only $F^+$ and transconjugants to grow for two more days. Fluorescence microscopy revealed transconjugants as yellow sectors that continued to grow after tetracycline application. (*B*) A mixed colony was grown for 4 days before tetracycline application, followed by two days of additional growth. The inner circle is the drop of the initial inoculant (1 μL). Once tetracycline is applied, only the $F^+_c$ cells and dark yellow transconjugants continue to grow. The tetracycline ring is outside the field of view. Scale bar is 1 mm.

A key feature of the spatial dynamics is the formation of monochromatic sectors composed of cells descending from either cyan or yellow fluorescent ancestors (Fig. 1*B*). Although a large number of individuals comprise the population, only a small number of individuals reproduce locally (i.e., at the nutrient-rich colony edge), leading to strong genetic drift. These demographic fluctuations reduce genetic diversity at the growing front and result in a single genotype reaching fixation locally and forming a small monochromatic domain. Over time, some of these domains grow while others disappear due to the random walk-like motion of the sector boundaries. Transconjugant sectors originate exclusively between a sector of $F^+$ and $F^-$ cells because conjugation can only occur when $F^+$ and $F^-$ cells are in physical contact.

**Limited spread of F plasmid in spatially structured populations**

The fate of the F plasmid depends on whether it can spread in a population. The rate of spread is determined by the fitness advantage or disadvantage conferred by the F plasmid and by the rate of plasmid transfer from $F^+$ to $F^-$ cells during conjugation. In the presence of tetracycline, the F plasmid confers a strong growth advantage and spreads in the population due to the increase in the number of $F^+$ cells, including transconjugants, relative to $F^-$ cells. In the absence of tetracycline, the F plasmid imposes a metabolic cost on its host (27); therefore, to survive it must spread through conjugation faster than $F^-$ cells outcompete $F^+$ cells. Previous experiments in the same system showed that the F plasmid spreads rapidly in exponentially growing well-mixed liquid cultures without tetracycline (5), in which the transconjugant fraction approaches 1 at a rate of 0.42 $h^{-1}$. In striking contrast, we saw that the fraction of cells with the plasmid stayed approximately constant or even slightly declined over time in a spatially structured population (Fig. 2; compare with Fig. 1 from Ref. (5), reprinted as Fig. S2). This qualitative change in the fate of the F plasmid shows that conjugation studied in well-mixed liquid cultures is a poor analog for conjugation in surface-associated colonies that more closely resemble natural populations.

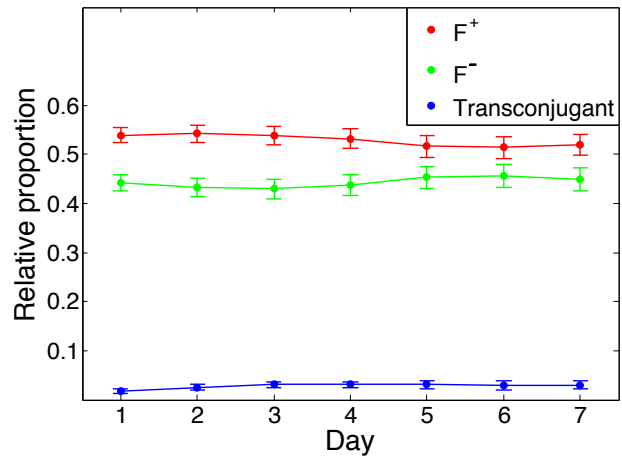

FIGURE 2 Dynamics of cell types in a conjugating spatially structured population. In contrast to the rapid ascension of transconjugants in well-mixed culture, transconjugants in spatial populations remain a small fraction of the population, as conjugation events are limited to the few boundaries between $F^+$ and $F^-$ sectors. The radial position of day $x$ was inferred as $x/7$ of total radial expansion during one week growth without tetracycline. Data shown are mean ± standard error (SE) of $n = 35$ colonies (1 μL inoculum).

**Accelerated loss of F plasmid in the presence of g3p**

To test whether molecular strategies for inhibiting conjugation were effective in the spatially structured population, we experimentally inhibited conjugation with a soluble form of the g3p protein of the M13 bacteriophage (g3p-N). At a protein concentration that decreases conjugation by 80% in liquid medium, the proportion of transconjugant cells at the colony front decreased by approximately 69% (from 5.2% to 1.6%, measured by circumference; Fig. 3). In addition the average number of transconjugant sectors decreased by approximately 53% (from 2.1 to 1.0, Fig. S3), confirming that g3p-N protein can indeed inhibit conjugation in surface-associated populations as well as in well-mixed populations.

**Model of conjugation in spatially structured populations**

We extended the one-dimensional 'stepping-stone' model of colony expansion to include conjugation. In the spirit of the Moran model (25) of evolution at constant population size, in which one individual reproduces and one individual dies during each time step, each generation consisted of a series of updates at which two cells were considered. As with previous modeling in this framework (23), these cells could exchange positions during a migration event, or one cell could die while the other divides during a reproduction event. In addition, here we introduced the possibility that one cell could transfer a plasmid to another during a conjugation event. The probabilities of these events were parameterized by the migration rate $m$, fitness cost of the F plasmid $s$, and conjugation rate $r$. The strength of genetic drift was controlled by $N$, the population size of each deme. The simulations were initialized by populating the demes with $F^+$ and $F^-$ cells drawn with equal probability. This procedure was similar to the well-mixed initial conditions in the experiments.

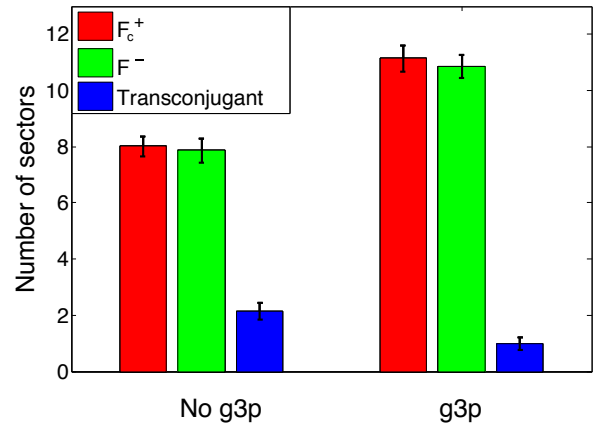

FIGURE 3 Cell types with and without conjugation inhibition by g3p-N. Circumference proportion of each cell type after 4 days of growth for 1:1 $F^+$:$F^-$ colonies (1 μL inoculum). The average proportion of transconjugants decreased threefold from 5.2% ± 1.0% to 1.6% ± 0.4% with addition of g3p. Data shown are mean ± SE of $n$ = 44 colonies without g3p-N and $n$ = 28 colonies with g3p-N.

Our simple model (Fig. 4*A-B*) qualitatively captured the experimentally observed formation of sectors and the appearance of transconjugant sectors (Fig. 4*C*). As in the experiments, we found that transconjugant sectors appeared between sectors of $F^+$ and $F^-$ cells, and that the number of transconjugant cells was smaller in spatial populations compared to well-mixed populations, suggesting that spatial structure could at least partially explain the stark difference in the fate of the F plasmid between liquid cultures and surface-associated colonies.

**Quantification of genetic drift and migration**

Genetic demixing, the most prominent feature of evolutionary dynamics in bacterial colonies, is controlled by the strength of genetic drift and migration. We quantified migration by $D_s$, the effective diffusion constant of sector boundaries, and genetic drift by $D_g$, the inverse of the product of the effective population density and the generation time. Here we followed the approach that has been previously applied to non-conjugating surface-associated microbial populations (24). For simplicity of the analysis, we performed experiments with two $F^-$ strains with different fluorescent colors since this avoids the complications of both the fitness cost of the F plasmid and conjugation.

We confirmed that experimental data indeed satisfied Eq. S3 and found $\frac{D_s}{v_\parallel} = 32 \ \mu m$ (Fig. 5*A*). Note that the expansion velocity $v_\parallel = 0.4$ mm/day and initial sector boundary position $R_i$ (which varied from colony to colony and boundary to boundary) were measured directly (Methods; Fig. S1). $D_s$ was also measured from simulations according to Eq. S4 (Table S1 and Figs. S4-6).

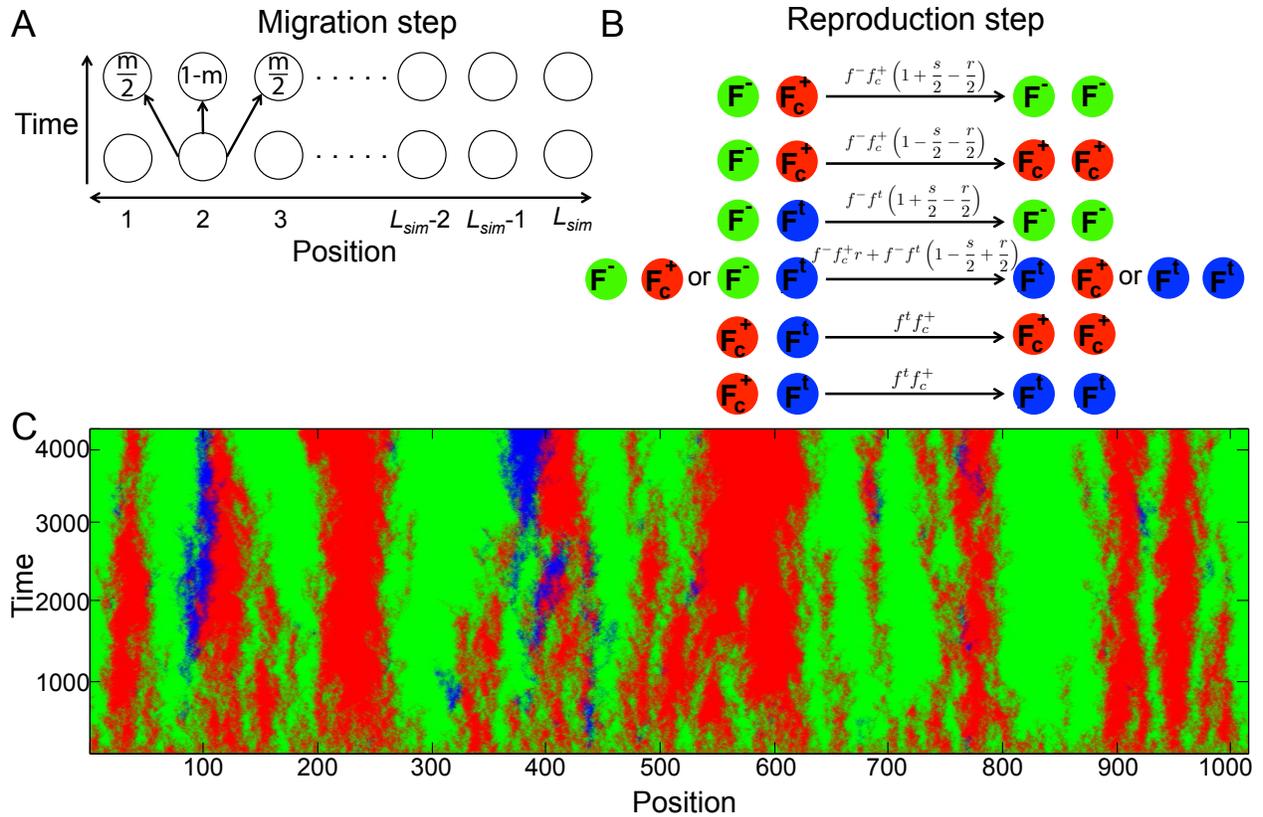

FIGURE 4 Simulation of conjugation during surface-associated growth. (*A*) Overview of the simulation: growth of the colony's population front outward over time is modeled by $L_{sim}$ demes with $N$ individuals each (indexed linearly with periodic boundary conditions). At the end of a generation, each individual migrates to either adjacent deme with probability $m/2$. (*B*) Each generation, all individuals are sequentially selected and undergo birth and death, which include selection (*s*), and conjugation (*r*) according the transition probabilities per generation (Eqs. 1-6) and the availability of interacting partners within the same deme. The probabilities in panel *B* do not sum up to one because some events do not change the composition of the population and therefore are not shown. (*C*) Simulated expansion shows good qualitative agreement with the experiments. This visualization with indexed deme position on the *x*-axis and generation number on the *y*-axis mimics experiments with $F_c^+$ cells (shown as red here), $F^-$ (green), and transconjugants (blue). Parameters correspond to the $N = 100$, $mN = 5$ simulation set in Table S4.

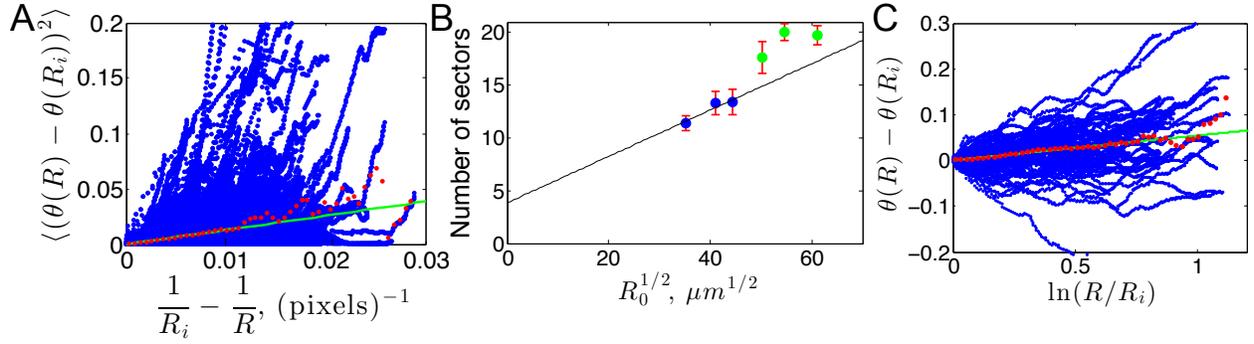

FIGURE 5 Quantification of sector patterns. (*A*) The diffusion of sector boundaries between differently labeled F⁻ cells of equal fitness was tracked in colonies grown for 18 days. We plotted 1373 individual sector boundaries as a function of radial position $R$ and initial boundary position $R_i$ in blue. The data was then split into 50 bins of equal length and averaged (red dots). The least squares line constrained through the origin as predicted by Eq. S3 was fit to the first 20 bins, which were not affected by sampling noise. The slope of the fit was used to estimate $\frac{D_s}{v_\parallel}$. Pixel size is 50 μm. (*B*) Number of sectors vs. $R_0^{1/2}$. The initial radii for the 1, 2, 3, 6, 10, and 15 μL colonies were averaged separately and the mean number of sectors ± SE was calculated. For the three largest drop sizes (green), there were a number of very small sectors that would likely be annihilated if grown for more time, but further growth would likely have suffered from nutrient deficiency and dehydrating conditions, so the line of best fit was calculated from the 1, 2, and 3 μL drop sizes (blue) constrained to the slope predicted by Eq. S5 with the previously calculated value of $\frac{D_s}{v_\parallel}$. The *y*-intercepts from panel *B* and Fig. S7, a second set of colonies, were averaged to 4, providing an estimate of $\frac{D_g}{v_\parallel}$. (*C*) The fitness difference between F⁺ and F⁻ cells in the absence of tetracycline was measured by plotting the sector boundary deviation as a function of the logarithm of the radial position. Boundaries were defined relative to F⁻ sectors, and one boundary of each sector was reflected so that all of them twist in the same direction. Most of the 76 sector edges grow outward, corresponding to positive $\frac{v_\perp}{v_\parallel}$. The trajectories were split into 50 bins of equal length to obtain the average behavior in red. The least squares line constrained to pass through the origin as predicted by Eq. S7 was fit to the first 15 bins (unaffected by sampling noise) to yield $\frac{v_\perp}{v_\parallel}$.

The strength of genetic drift was measured experimentally according to Eq. S5, which predicts that the number of sectors grows as the square root of the initial colony radius. We varied $R_0$ by inoculating the colonies with different amounts of well-mixed liquid culture and confirmed the square root dependence (Figs. 5B and S7). Then, $\frac{D_g}{v_\parallel} = 0.785$ was estimated by fitting Eq. S5 to the data using our previous estimate of $D_s$. The fit to the simulation data also yields an estimate of $D_g$, as described in the Methods (Table S2 and Figs. S4-6).

**Fitness cost of the F plasmid**

We observed slight boundary bending in the experimental data, suggesting a fitness cost to the F plasmid. This is consistent with Eq. S7 (Fig. 5C) and we thus estimated $\frac{v_\perp}{v_\parallel} = 0.054$. An analogous procedure was used in simulations (Table S3).

**Quantification of conjugation**

In both experiments and simulations, we quantified conjugation by the fraction of the colony circumference occupied by the transconjugant cells (5.2% and 1.6% for 1 $\mu$L inoculum grown for 4 days without and with g3p-N, respectively, Fig. 3). We emphasize that the number of transconjugant sectors was not used to parameterize the model, but was used to check the agreement between experiments and simulations.

**Match between experiments and modeling accurately quantifies conjugation**

Experimental data were used to parameterize the model as described in the Methods. The results of this matching are summarized in Table 2 and Tables S4 and S5, and visualizations are presented in Fig. S8. Note that we report conjugation rate per unit of time using the relation $r_{exp} = \frac{r_{sim}T_{sim}}{T_{exp}}$ because $rT$ is a dimensionless invariant.

We checked whether the parameterized model was able to correctly predict additional biological results, which had not been included during parameterization (Table 2, Consistency checks). First, we found that the model parameterization successfully captured the inhibition of conjugation by g3p-N protein (Table 2). Using experimental data from the surface-associated populations, the modeling estimated that addition of 10 nM g3p-N reduces the effective conjugation rate by more than a factor of three, from $7.6\times10^{-3}$ h$^{-1}$ to $2.4\times10^{-3}$ h$^{-1}$. This reduction is similar to previously reported results in liquid culture (5), in which 10 nM g3p-N reduced the rate of transconjugant spread by a factor of 2.5. Second, another success of the model is the accurate prediction of the number of transconjugant sectors, both with and without g3p-N protein (Tables 2, S6-7), which was not used in parameterizing the model.

Using our model, we explored how the fate of conjugative plasmid depends on the selective coefficient and conjugation rate in spatial populations (Figs. 6 and S9). As expected, higher conjugation rates facilitated plasmid spread while higher fitness costs inhibited it. Our model predicts that the conjugative plasmid is lost when $s>r$ because the fraction of plasmid free F$^-$ cells depends only on ($s-r$); see Eq. 7. We expect this result to hold in more complex communities as long as both competition and conjugation occur locally (e.g. along the sector boundaries).

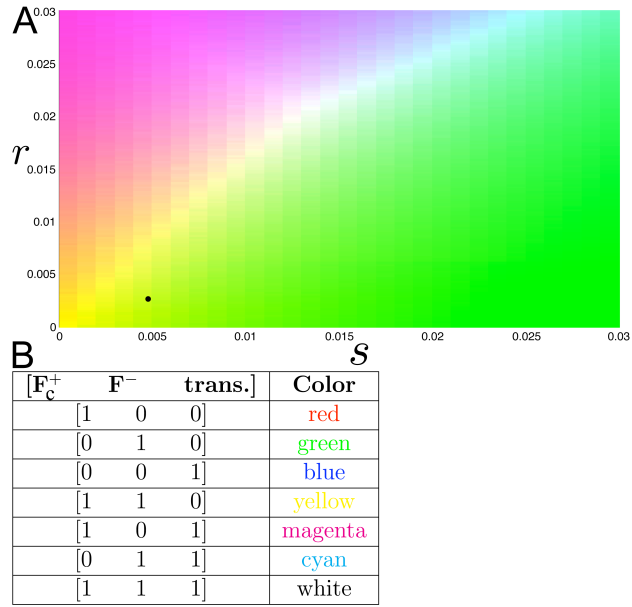

| $[F^+_c$ | $F^-$ | trans.] | Color |
|---|---|---|---|
| [1 | 0 | 0] | red |
| [0 | 1 | 0] | green |
| [0 | 0 | 1] | blue |
| [1 | 1 | 0] | yellow |
| [1 | 0 | 1] | magenta |
| [0 | 1 | 1] | cyan |
| [1 | 1 | 1] | white |

FIGURE 6 Simulation outcome as a function of conjugation ($r$) and selection ($s$). (*A*) Simulations of $N = 30$, $mN = 1$ were conducted over a range of 150 $r$ values from 0 to 0.03 and 30 $s$ values from 0 to 0.03. The values that correspond to 4 days of experimental growth without g3p are $s = 4.8\times10^{-3}$ and $r = 2.63\times10^{-3}$ (5.2% transconjugants) as marked by the black dot. The coloring represents the total fraction of individuals of each type over the entire population at $T_{sim} = 252$ generations. The green regions represent almost 100% $F^-$ individuals as expected for high $s$. The white regions represent an equal mix of $F^+_c$, $F^-$, and transconjugants. The yellow regions represent 50% $F^+$ and 50% $F^-$ corresponding to low $s$ and $r$ in which few population-changing events occur. (*B*) Shading of the relative population composition for each simulation in panel *A*.

Many natural conjugative plasmids might be living close to an extinction threshold; otherwise they would either spread rapidly or go extinct. Although the threshold for plasmid spread ($r=s$) is the same in both spatial and well-mixed populations, their evolution is quite different. When $r \approx s$, $F^+$ cells in well-mixed populations are primarily transconjugants because original $F^+$ cells are outcompeted. Therefore, genetic background of initially $F^+$ cells is lost and the conjugative plasmid spreads primarily via horizontal transmission. In contrast, spatial populations would have a much smaller fraction of transconjugants under similar conditions, which is evident from the lack of transconjugant-dominated blue along the $r=s$ diagonal in Fig. 6. In spatial populations, competition and conjugation would occur near the boundaries and the bulk of the original $F^+$ cells will be shielded from competition with $F^-$ cells. As a result, the genetic background of initially $F^+$ cells is preserved and vertical transmission is a primary mechanism of plasmid persistence. Such differences are likely to play an important role in the evolution of both bacteria and their conjugative plasmids.

**DISCUSSION**

We analyzed conjugation of an F plasmid carrying tetracycline resistance in bacterial colonies growing on an agar surface. The genetic history of the colony could be visualized in the fluorescence pattern, which distinguished between donor ($F^+_c$), potential recipient ($F^-$), and transconjugant cells. As expected, conjugation events occurred only at boundary zones between $F^+$ and $F^-$ cells. However, in spatially structured populations, the number of boundary zones was surprisingly small, even when the populations were initially well-mixed, because strong genetic drift at the colony edge led to stochastic separation of different strains in space, i.e., genetic demixing. As a result, the F plasmid failed to spread in the population, in sharp contrast to liquid culture, where $F^-$ and $F^+$ cells easily come in contact and the F plasmid spreads through the population like an epidemic, approaching complete conversion to the $F^+$ genotype (5).

We have extended previous models of conjugation to include spatial structure and genetic demixing during range expansions in bacterial populations growing on a surface. Stochastic simulations were necessary to capture the important element of genetic drift in the spatially structured environment, as drift determines the availability of zones in which $F^+$ and $F^-$ cells can contact one another. Our parameterization method using invariants meant that experimental invariants determined simulation invariants with no additional degrees of freedom. Our model makes the simplifying assumptions that conjugation and competition are first-order reactions, and that both donor and transconjugant cells are immediately capable of future conjugation after a conjugation event, although a refractory period is known to follow conjugation events (28, 29). We also neglected the circular geometry of the colonies and used discrete demes to model a continuous population. Nevertheless, our simulations based on a one-dimensional stepping-stone model accurately describe experimental outcomes and could be used to estimate the effective conjugation rates from the data.

As noted by others, previous measurements of plasmid transfer rates in surface associated populations have been difficult to interpret (15). The commonly used "endpoint method" (30) provides a reliable estimate for homogeneously-mixed populations, but it has also been applied to bacterial populations with spatial structure (31-33), for which assumptions of the method are violated. While any pair of donor and recipient cells can conjugate in a well-mixed population, conjugation is restricted to spatial neighbors in a surface-associated population, where the distribution of donors and recipients on very small spatial scales determines the rate of production of transconjugant cells (31, 33, 34). The approach taken here incorporates this crucial fact by treating the population as a set of demes connected by nearest-neighbor migration and restricting plasmid transfer only to cells present in the same deme. The important role of stochastic fluctuations, which cause genetic demixing and a qualitative change in conjugation behavior in the population, is also neglected by the endpoint method. Our method includes these effects, and the resulting estimate of the effective conjugation rate could allow quantitative comparisons of transfer efficiencies across different plasmids and environments (13, 35, 36). Thus, visualization and analysis of expanding and conjugating microbial populations on surfaces is analytically, experimentally and computationally tractable, enabling in-depth study of the dynamics of gene transfer.

**CONCLUSION**

While the F plasmid is the most well-studied conjugative system, many others exist and are medically important in the spread of antibiotic resistance (37). Such systems should be studied in realistic settings, such as surfaces and biofilms, where stochastic effects heavily influence the genetic outcome. Our findings suggest that strategies to enforce spatial structuring could reduce the spread of undesirable genes to new organisms even though the donor cells themselves may continue to reproduce and constitute a large fraction of the microbial population. Such ecological strategies are complementary to attempts to block conjugation at a molecular level. Indeed, we found that spatial structure from surface growth combined with an inhibitor of conjugation produces a multiplicative decrease in the conjugation rate. Such anti-conjugation strategies may be worthy of further study as resistance to antibiotics becomes increasingly widespread among pathogenic bacteria.

**ACKNOWLEDGMENTS**

The authors thank David Nelson for advice and Erin O'Shea for use of equipment. P.D.F. was supported by HCRP and PRISE fellowships at Harvard College. K.S.K. was supported by an MIT Pappalardo Fellowship in Physics. J.I.J. was a Foundational Questions in Evolutionary Biology fellow at Harvard University sponsored by the John Templeton Foundation. I.A.C. was a Bauer Fellow at Harvard University and is a Simons Investigator (grant 290356 from the Simons Foundation). This work was also supported by NIH grant GM068763 to the National Centers of Systems Biology and grant RFP-12-05 from the Foundational Questions in Evolutionary Biology Fund.

**TABLES**

| Parameter | Experimental data | Simulation data |
|---|---|---|
| Migration ($D_s$) | Wandering of sector boundaries | Global heterozygosity (probability that two cells from the colony are the same type) |
| Genetic drift ($D_g$) | Number of surviving sectors | Local heterozygosity (probability that two cells from a deme are the same type) |
| Cost of the conjugative plasmid ($s$) | Bending of sector boundaries | Bending of sector boundaries |
| Spatial and temporal scales | Measured distance and time | Deme number and size |

TABLE 1 **Parameterization sources** Description of model parameters and their analogs in experimental and simulation data. Parameters were combined to calculate dimensionless invariants, as described in the Methods, to match experimental and simulation data.

| Quantity | Experimental measurement | Simulation quantity | Data utilized |
|---|---|---|---|
| **Time** | $T = 4$ days | $T_{sim} = 4295$ generations | Time of growth |
| **Length** | $L = 1.6$ cm | $L_{sim} = 1015$ demes | Average colony circumference |
| **Migration** | $\frac{D_s}{v_\parallel} = 32\ \mu m$ | $m = 0.05$/generation | Random walk of boundaries |
| **Genetic drift** | $\frac{D_g}{v_\parallel} = 0.785$ | $N = 100$ | Number of sectors |
| **Plasmid cost** | $\frac{v_\perp}{v_\parallel} = 0.054$ | $s = 4.5 \times 10^{-4}$ | Deterministic bending of boundaries |
| **Conjugation rate without g3p** | $r = 7.6 \times 10^{-3}$/hour | $r = 1.7 \times 10^{-4}$, unitless | Transconjugant circumference proportion |
| **Conjugation rate with g3p** | $r = 2.4 \times 10^{-3}$/hour | $r = 5.5 \times 10^{-5}$, unitless | Transconjugant circumference proportion |
| **Consistency check: Transconjugant sectors without g3p** | $2.14 \pm 0.31$ | $2.05 \pm 0.30$ | Number of transconjugant sectors |
| **Consistency check: Transconjugant sectors with g3p** | $1.00 \pm 0.22$ | $1.00 \pm 0.13$ | Number of transconjugant sectors |

TABLE 2 **Quantification of conjugation in experiments and simulations** All quantities except the conjugation rate were measured in experiments and their simulation counterparts were fit to match the experimental setup. Then, the simulation conjugation rate was fit to the observed transconjugant circumference proportion to determine the experimental conjugation rate. As a consistency check, we observed that the number of sectors in simulations agreed with experimental conditions without and with g3p-N protein even though these values were not used to match the conjugation rate.

Supporting Material for:

Genetic drift suppresses bacterial conjugation in spatially structured populations

Peter D. Freese,[†] Kirill S. Korolev,[†‡§¶] José I. Jiménez,[†∥] and Irene A. Chen[†∥**][*]

[†]FAS Center for Systems Biology, Harvard University, Cambridge, Massachusetts; [‡]Department of Physics, Harvard University, Cambridge, Massachusetts; [§]Department of Physics, Massachusetts Institute of Technology, Cambridge, Massachusetts; [¶]Department of Physics and Program in Bioinformatics, Boston University, Boston, Massachusetts; [∥]Faculty of Health and Medical Sciences (University of Surrey, UK) and [**]Department of Chemistry and Biochemistry, Program in Biomolecular Sciences, University of California, Santa Barbara, California

*Correspondence: chen@chem.ucsb.edu

**Contents**

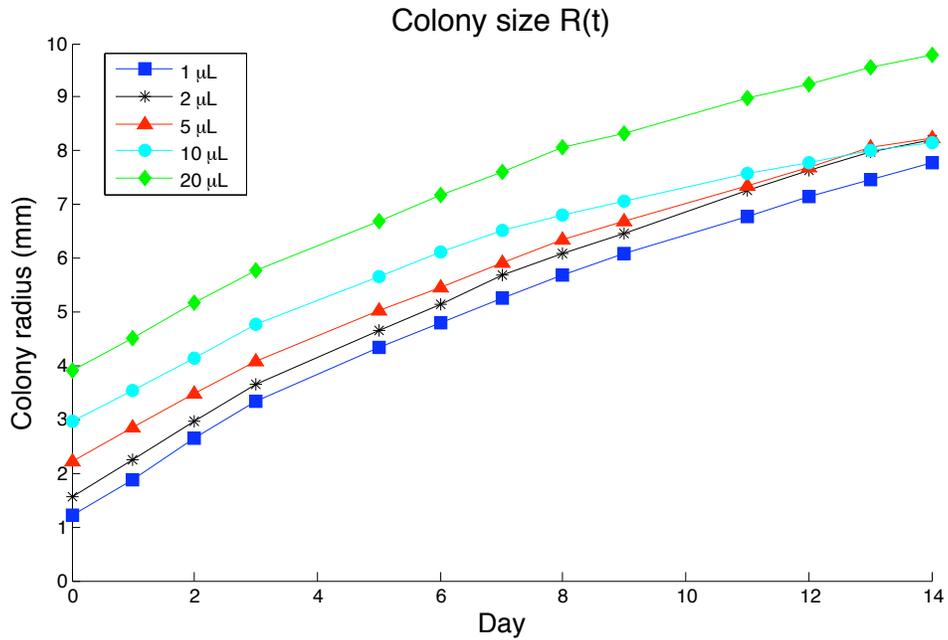

FIGURE S1 Average colony radius of *n* = 30 plates over 14 days, for several inoculum volumes. There is a gradual decrease in expansion velocity, but the expansion over the first four days used for most analyses here is well approximated by a constant velocity. We find that $v_\parallel \approx$ 0.4 mm/day = $4.6 \times 10^{-3}$ μm/second, and the average circumference for a 1 μL colony over 4 days of growth was 1.6 cm.

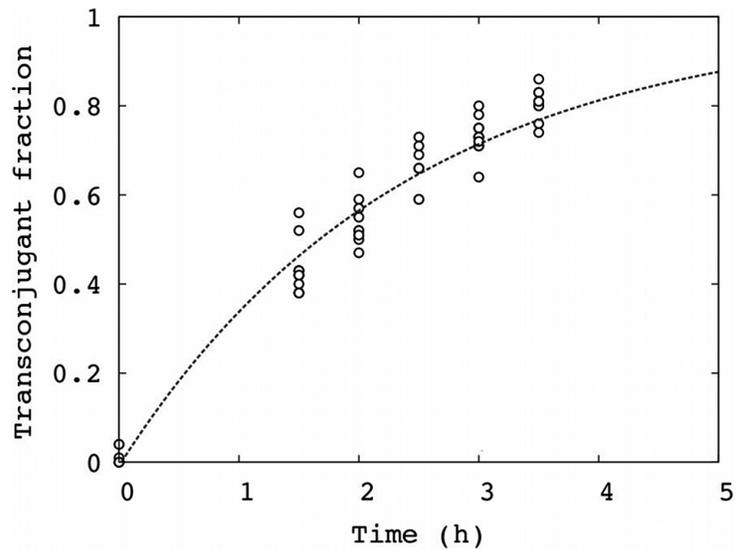

FIGURE S2 Reprinted Fig. 1 from Sup. Ref. (1), showing an increase in transconjugants from fraction 0 to approaching 1 exponentially in well-mixed $F^-/F^+$ culture in the absence of phage. In contrast, we found that in our spatially-structured environment the transconjugant fraction levels off at about 0.05 (Main Text Figs. 2, 3).

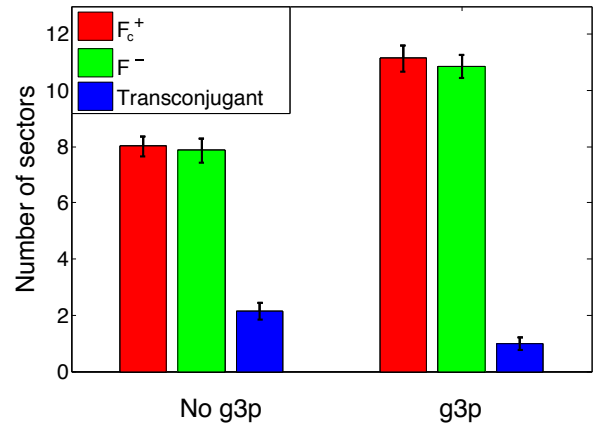

FIGURE S3 Number of sectors in experiments with and without conjugation inhibition. Data from the same colonies as Main Text Fig. 3. The mean number of transconjugant sectors ± SE, which decreased about twofold from 2.14 ± 0.31 to 1.00 ± 0.22. See also Tables S6 and S7. A possible explanation for the differences in the absolute number of sectors with and without g3p is that application of the liquid g3p mixture changed the agar surface tension after drying and therefore the contact angle of the inoculation, increasing $R_0$. Alternatively, because colonies were grown on separate days, differences could be due to variation in selected clones or ambient conditions.

## Methods: Modeling Details

The inference of $D_s$ and $D_g$ was done in pure F⁻ colonies without the complications of conjugation and fitness differences. For simulated linear expansions, the dynamics of two cell type frequencies $f$ and $1-f$ is described by a simple continuous theory in terms of the following stochastic partial differential equation from Supporting References (2, 3):

$$\frac{df}{dt} = D_s \frac{\partial^2 f}{\partial x^2} + \left(D_g f(1-f)\right)^{\frac{1}{2}} \Gamma(t,x), \tag{S1}$$

where the spatial diffusion constant $D_s = \frac{ma^2}{2\tau}$ is related to migration and the genetic diffusion constant $D_g = \frac{a}{N\tau}$ is related to genetic drift. The spatial coordinate is $x = la$, where $l$ indexes the demes, $a$ is the distance between demes, $\tau$ is the generation time, and $\Gamma$ is white zero mean Gaussian noise. For circular expansions, it is convenient to use polar coordinates around the center of the colony with initial radius $R_0$. Equation S1 is then replaced by:

$$\frac{\partial f}{\partial t} = \frac{D_s}{(R_0 + v_\parallel t)^2} \frac{\partial^2 f}{\partial \theta^2} + \left(D_g \frac{f(1-f)}{(R_0 + v_\parallel t)}\right)^{\frac{1}{2}} \Gamma(t,\theta), \tag{S2}$$

where $v_\parallel$ is radial expansion velocity (2, 3).

### Quantifying migration: calculation of diffusion constant $D_s$

The effective diffusion constant $D_s$ can be measured from the amount of boundary wandering around its mean position. If $\theta(R)$ is the angular position of the sector boundary, then its mean squared deviation is given in Sup. Ref. (2):

$$\langle (\theta(R) - \theta(R_i))^2 \rangle = \frac{2D_s}{v_\parallel} \left(\frac{1}{R_i} - \frac{1}{R}\right), \tag{S3}$$

where $R_i$ is the beginning of the sector boundary. This hyperbolic increase in the variance of boundary position is due to the circular nature of the expansion; as the colony grows, the angular diffusion vanishes because the same interval in $x$ space corresponds to a continually smaller interval in $\theta$ space. We used Eq. S3 to estimate $D_s$ from the experiments.

To obtain a large number of sector boundaries to fit to Eq. S3, an equal mixture of cells was inoculated onto 67 plates with drop sizes ranging from 1 µL to 15 µL. To minimize fitness differences that would complicate analysis, the inoculant consisted of F⁻ cells differing only in their fluorescent protein color (cyan or yellow). Colonies were grown for 18 days to obtain long boundaries, yielding 1373 sector boundaries. The $x$-axis was separated into 50 bins of equal length, and all points within each bin were averaged to obtain $\langle(\theta(R) - \theta(R_i))^2\rangle$. The least squares error line was fit to the first 20 binned points, which were covered by the majority of sector boundaries. The other bins had large fluctuations due to small number of sector boundaries contributing to the average. We also constrained the fit to go through the origin, as required by Eq. S3. This procedure yielded $\frac{D_s}{v_\parallel} = 32 \mu m$ (Fig. 5A).

To measure $D_s$ in simulations, we used the global heterozygosity, the probability of sampling two cell types regardless of their position in the population. This quantity $\mathcal{H}(t)$ is given in Sup. Ref. (2) and decreases for intermediate times as:

$$\mathcal{H}(t) = \frac{1}{2} - \frac{1}{L}\left(\frac{8D_s t}{\pi}\right)^{1/2}, \tag{S4}$$

enabling measurement of $D_s$. This decrease is given by the fluctuations in the sizes of monochromatic sectors, which behave approximately as random walks (thus $(D_s t)^{1/2}$ scaling).

**Quantifying genetic drift: calculation of diffusion constant $D_g$**

The strength of genetic drift $D_g$ was estimated from the number of surviving sectors, which eventually becomes time-independent because angular diffusion of sector boundaries and their coalescence vanish due to the radial expansion of the colony. The number of surviving sectors ($N_s$) grows as the square root of the initial colony radius as given by the following equation in Sup. Ref. (2):

$$N_s(t \to \infty) = \frac{\pi v_\parallel}{D_g} + \left(\frac{\pi R_0 v_\parallel}{2 D_s}\right)^{1/2}. \tag{S5}$$

The second term on the right hand side describes the number of sectors in the limit of immediate sector formation followed by coalescence of sector boundaries. The first term is the additional number of sectors due to the finite time necessary to form sectors, which strongly depends on the strength of genetic drift $D_g$.

We used Eq. S5 to estimate $D_g$. The fit of experimental data to Eq. S5 is shown for two sets of colonies in Fig. 5B (set 1) and Fig. S7 (set 2). The y-intercepts from Fig. 5B and Fig. S7 averaged to 4, yielding $\frac{D_g}{v_\parallel} = 0.785$.

In simulations, the relative proportions of different cells in each deme were known, which enabled a simpler and more accurate method of estimating $D_g$. We computed the average local heterozygosity, i.e., the probability of sampling two different cell types from the same deme (with replacement). This quantity decays to zero as $\left(\frac{2 D_s}{\pi D_g^2 t}\right)^{\frac{1}{2}}$ at long times (2, 3). This inverse square root scaling is again related to the random walk-like motion of sector boundaries.

**Quantifying the fitness cost of the F plasmid**

In the absence of tetracycline, the F plasmid is metabolically costly to the cell. To estimate the magnitude of this fitness cost, we examined the behavior of sector boundaries between F$^+$ and F$^-$ cells. For the circular geometry of microbial colonies, sector boundary motion is described by the following stochastic differential equation from Sup. Ref. (3):

$$\frac{d}{dr}\theta(r) = \frac{v_\perp}{v_\parallel r} + \left(\frac{2D_s}{v_\parallel r^2}\right)^{1/2} \Gamma(r), \tag{S6}$$

where $v_\perp$ is the velocity of the sector boundary perpendicular to the direction of the expansion through which the fitter strain invades the less fit strain. The $\frac{v_\perp}{v_\parallel}$ ratio reflects the fitness cost of maintaining and replicating the F plasmid in the absence of tetracycline. This fitness cost is manifested by the expansion of F$^-$ sectors at the expense of F$^+$ sectors.

By averaging and integrating Eq. S6, we obtain the mean sector boundary as given in Sup. Ref. (4):

$$\langle\theta(R)\rangle = \theta(R_i) + \frac{v_\perp}{v_\parallel} \ln\left(\frac{R}{R_i}\right), \tag{S7}$$

We used Eq. S7 to estimate $\frac{v_\perp}{v_\parallel}$ from experimental data.

We only used long boundaries that were far apart so as not annihilate. To increase the number of such boundaries, F$^-$:F$^+$ ratios from 1:8 to 1:64 were prepared and inoculated in drop sizes 1 μL-20 μL to get a range of $R_i$. Over all of the plates, 38 F$^-$ sectors resulted in 38 clockwise-bending and 38 counterclockwise-bending F$^+$:F$^-$ edges. Half of the edges were reflected so that $\frac{v_\perp}{v_\parallel}$ had the same sign for all 76 edges, and their sector boundaries were analyzed using Eq. S7. The *x*-axis was split into 50 bins of equal length and points within each bin were averaged to obtain $\langle\theta(R)\rangle$. The line of best fit was calculated according to the first 15 bins, which were covered by a majority of sector boundaries and did not suffer from noise. The fit was constrained to go through the origin and yielded $\frac{v_\perp}{v_\parallel} = 0.054$ (Fig. 5*C*).

In simulations, $v_\perp$ was estimated using the same procedure, but, instead of logarithmic spirals, we fit straight lines to sector boundaries, as appropriate for linear geometry.

## Data: Simulation Details

In our experiments, the change in the circumference was only about a factor of four, suggesting that taking $L_{exp}$ to be twice the initial circumference of the colony could be a good approximation, especially given that most of the dynamics occurs early on during the expansion (2).

In the absence of conjugation and fitness differences between the strains, there are four quantities that characterize simulations: the population size of each deme $N$ (a measure of genetic drift), the migration rate $m$ (a measure of mixing), the number of simulated generations $T_{sim}$ (a measure of time), and the number of demes $L_{sim}$ (a measure of length). Naively, one might think that all four parameters can be estimated from the experimental data such as the image shown in Fig. 1B. This turns out not be the case because the parameters are not independent of one another, and one can chose any two of the four parameters freely and then use the remaining two parameters to fit the data; see also Sup. Ref. (2). One reason for this freedom is that the subdivision of the population into demes can be arbitrary, and any subdivision produces similar sectoring patterns on spatial scales larger than the size of a deme, provided $m$ and $N$ are chosen appropriately. For example, combining every two nearby demes into a larger deme simply maps $N$ into $2N$, $L_{sim}$ into $L_{sim}/2$, and $m$ into $m/2$ (only half of the migrants escape the double-sized deme) and leaves the spatial scales unaltered. Similarly, the choice of the generation time is arbitrary because smaller populations experience the same amount of genetic drift in a shorter time as larger populations in a longer time. This creates another degree of freedom in choosing the parameters of the simulations.

One way to estimate all four parameters is to use the actual generation time of bacteria and use $N = 1$ considering each bacterium as its own deme. This and similar approaches have two drawbacks. First, generation times vary at the front; presumably bacteria closer to the front have access to more nutrients and divide faster. Similarly, there are variations in bacterial density and movement patterns at the front which invalidate the assumption of invariant $N = 1$. Second, simulating individual bacteria on the spatial scales of interest to us here requires infeasible computational power. In contrast, our approach of using an effective one-dimensional description is computationally tractable and quantitatively captures the observed spatial patterns as has been demonstrated (2, 3).

For computational efficiency, we chose to run simulations for three sets of ($N$, $mN$): ($N = 30$, $mN = 1$), ($N = 30$, $mN = 10$), and ($N = 100$, $mN = 5$) using values of $T_{sim}$ and $L_{sim}$ estimated from the data as we explain below. Our results were essentially the same for all three set of parameters further supporting this modeling approach.

The values of $T_{sim}$ and $L_{sim}$ that match the experiments were determined by applying a continuous theory of spatial genetic demixing which has previously been used to model range expansions in bacterial colonies (2). We first found the values of the parameters in the continuous theory that described our experiments and then fit the simulation parameters to give the same continuous theory values as those from our experiments.

For 1 µL colonies grown for $T_{exp} = 4$ days, the average growth rate and colony circumference are $v_\parallel = 4.6 \times 10^{-3}$ µm/sec and $L_{exp} = 1.6$ cm, respectively (Fig. S1). Using the experimental spatial and genetic diffusion constants previously calculated yields invariant quantities of $Inv_1 = 3 \times 10^{-2}$ and $Inv_2 = 2.5 \times 10^{-3}$ (Main Text Eqs. 10 and 11, respectively).

Solving for $T_{sim}$ and $L_{sim}$ in the simulation version of Main Text Eqs. 10 and 11 yields:

$$T_{sim} = \frac{D_s}{D_g^2} \frac{1}{Inv_1}, \qquad (S8)$$

$$L_{sim} = \frac{D_s}{D_g} \frac{1}{Inv_2}. \qquad (S9)$$

Substituting in the invariant quantities of $Inv_1 = 3\times10^{-2}$ and $Inv_2 = 2.5\times10^{-3}$ and the diffusion constant values calculated below yields $T_{sim}$ and $L_{sim}$ for each simulation set (Table S4).

## Calculation of conjugation rate *r*

### Strategy for determination of conjugation rate *r*

The simulation diffusion constants $D_s$ and $D_g$ must be calculated to determine $T_{sim}$ and $L_{sim}$ that match the experiments. The overall simulation strategy is, for each (*N*, *mN*) simulation set:

1. Determine simulation $D_s$ and $D_g$ that correspond to experimental $D_s$ and $D_g$.
2. Calculate the $T_{sim}$, $L_{sim}$ and the selective coefficient *s* that correspond to these $D_s$ and $D_g$ according to Eqs. S8 and S9 and the experimentally calculated invariants.
3. Run simulations for $T_{sim}$ generations with $L_{sim}$ demes and the selective coefficient *s* for a range of candidate *r* values. Match the simulation transconjugant circumference proportion to experiments to infer the correct conjugation rate *r*.

**Calculation of diffusion constants $D_s$ and $D_g$**

For linear geometry, the probability of sampling two different cell types regardless of their spatial separation $x$ is the global heterozygosity (2):

$$\mathcal{H}(t) = \frac{1}{L}\int_0^L H(t,x)dx, \tag{S10}$$

which, for $r = s = 0$, obeys:

$$2F_0(1 - F_0) - \mathcal{H}(t) = \frac{1}{L}\left(\frac{8D_s t}{\pi}\right)^{1/2} + O\left(\frac{D_s}{D_g L}\right), \tag{S11}$$

assuming $\frac{8D_s}{D_g^2} \ll t \ll \frac{L^2}{D_s}$. Equation S4 is derived from Eq. S11 with initial cell type frequency $F_0 = \frac{1}{2}$. Equation S11 describes a linear relationship between $2F_0(1 - F_0) - \mathcal{H}(t)$ and $t^{1/2}$ with slope $\frac{1}{L}\left(\frac{8D_s}{\pi}\right)^{1/2}$, allowing $D_s$ to be computed.

For each of the three simulation sets, 25 realizations were run with $L_{sim} = 2000$ demes for $t = 1{,}500{,}000$ generations. The 25 realizations per simulation set were used to calculate the global heterozygosity at generation $t$:

$$H(t) = \frac{1}{L}\sum_{i=1}^{L} H_i(t), \tag{S12}$$

where $H_i(t)$ is the local well-mixed heterozygosity within deme $i$. The global heterozygosity for each generation was then averaged over the 25 realizations. Because the relationship is linear only for $\frac{8D_s}{D_g^2} \ll t \ll \frac{L^2}{D_s}$, the lines of fit in Table S1 correspond to times in this range for expected parameter values.

| $2F_0(1 - F_0) - \mathcal{H}(t)$ vs. $t^{1/2}$ to compute $D_s$ | | |
|---|---|---|
| **Simulation set** | **Line of best fit** | $D_s$ |
| $N = 30$, $mN = 1$ | $y = 0.000145x - 0.0075$ | 0.033 |
| $N = 30$, $mN = 10$ | $y = 0.000344x - 0.038$ | 0.18 |
| $N = 100$, $mN = 5$ | $y = 0.0001755x - 0.025$ | 0.048 |

TABLE S1 **The line of best fit of $2F_0(1 - F_0) - \mathcal{H}(t)$ vs. $t^{1/2}$ is used to infer $D_s$ for each simulation set** See Figs. S4-6(*B*) for the full plots.

The decrease in heterozygosity within a deme for times $t \geq 8D_s/D_g^2$ is given in Sup. Ref. (2):

$$H(t) = \left(\frac{\pi D_g^2 t}{2D_s}\right)^{-\frac{1}{2}} + O\left(t^{-\frac{3}{2}}\right). \tag{S13}$$

For large times, $t^{-1/2} \to 0$ so $H(t) \to 0$ and one of the cell types reaches fixation locally. Thus, the spatial model is consistent with experiments because it predicts the formation of sectors in which one of the cell types is locally fixed. Equation S13 yields a linear relationship between $H(t)$ and $t^{-1/2}$ through the origin with slope $\left(\frac{\pi D_g^2}{2D_s}\right)^{-\frac{1}{2}}$, allowing $D_s/D_g^2$ to be computed.

For each of the three simulation sets, the same 25 realizations from Table S1 were used. At each generation, the heterozygosity within the deme was calculated and averaged over all 2000 demes; the heterozygosity for each time point was then averaged over all 25 realizations. Because Eq. S13 is valid for $t \geq 8D_s/D_g^2$, the lines of fit in Table S2 correspond to $t^{-1/2} \leq 0.007$.

| | $H(t)$ vs. $t^{-1/2}$ to compute $D_s/D_g^2$ | |
|---|---|---|
| **Simulation set** | **Line of best fit** | $D_s/D_g^2$ |
| $N = 30$, $mN = 1$ | $y = 2.194x + 0.00023$ | 7.516 |
| $N = 30$, $mN = 10$ | $y = 4.216x + 0.0012$ | 27.92 |
| $N = 100$, $mN = 5$ | $y = 9.043x + 0.0007$ | 128.45 |

TABLE S2 **The line of best fit of $H(t)$ vs. $t^{-1/2}$ for $t^{-1/2} \leq 0.007$ is used to infer $D_s/D_g^2$ for each simulation set** See Figs. S4-6(*A*) for the full plots.

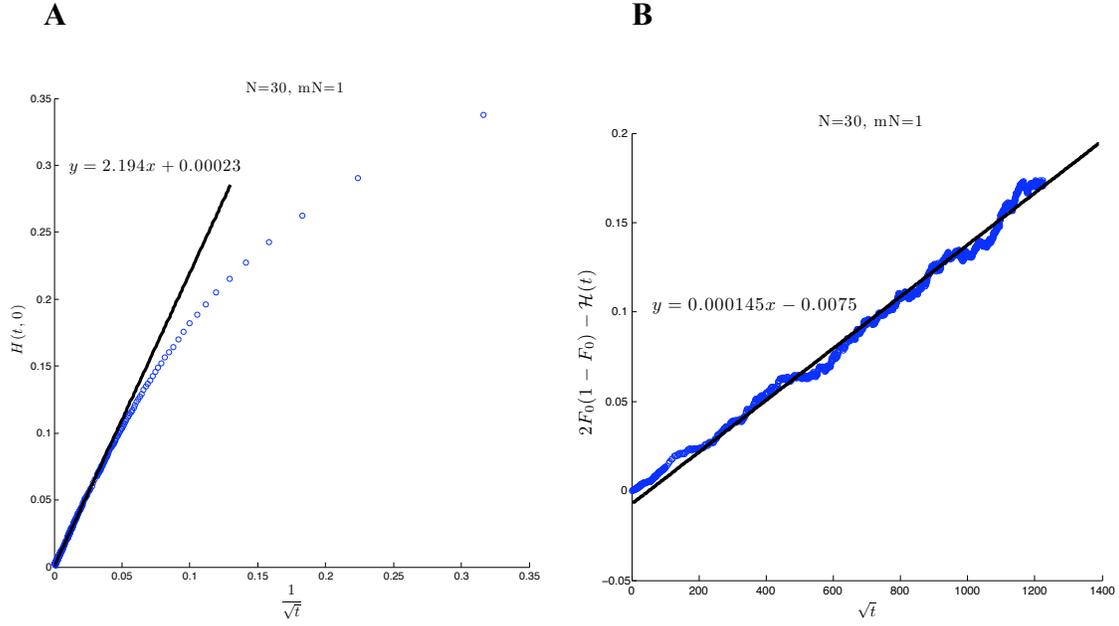

FIGURE S4 Parameter fitting of ($N = 30$, $mN = 1$) simulation set. (*A*) The slope of the best fit line for $t^{-1/2} \leq 0.007$ was used to estimate $D_s/D_g^2$ according to Eq. S13. (*B*) The slope of the line for intermediate $t^{1/2}$ was used to calculate $D_s$ according to Eq. S11.

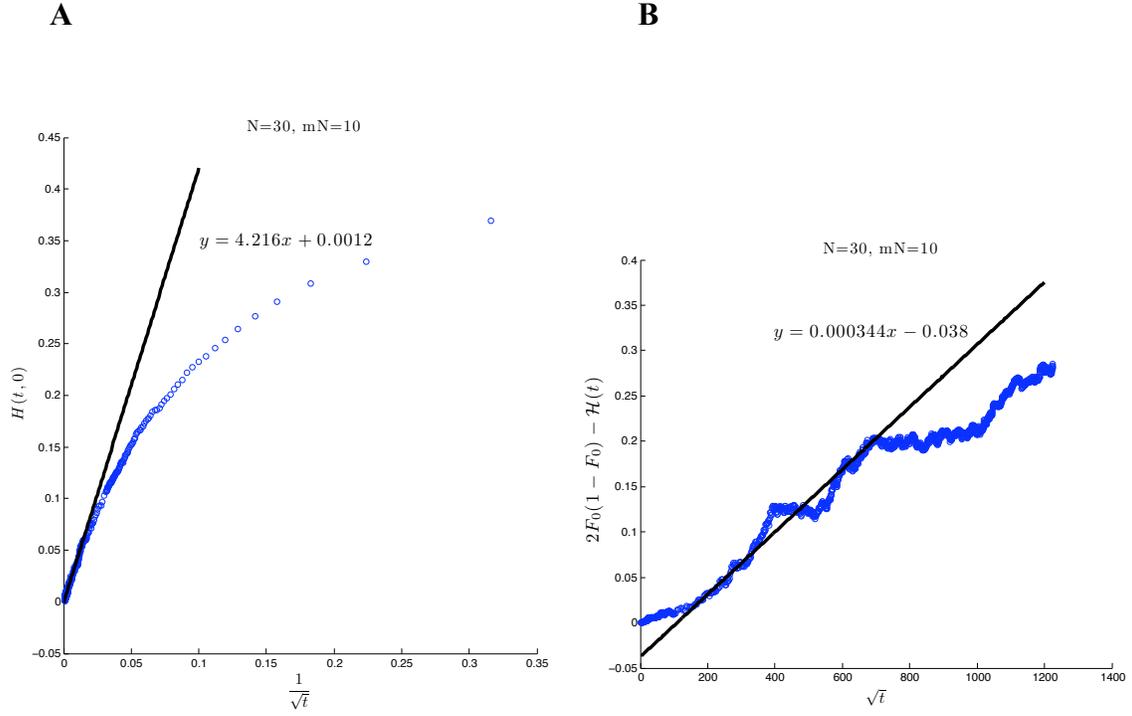

FIGURE S5 Parameter fitting of ($N = 30$, $mN = 10$) simulation set. (*A*) The slope of the best fit line for $t^{-1/2} \leq 0.007$ was used to estimate $D_s/D_g^2$ according to Eq. S13. (*B*) The slope of the line for intermediate $t^{1/2}$ was used to calculate $D_s$ according to Eq. S11.

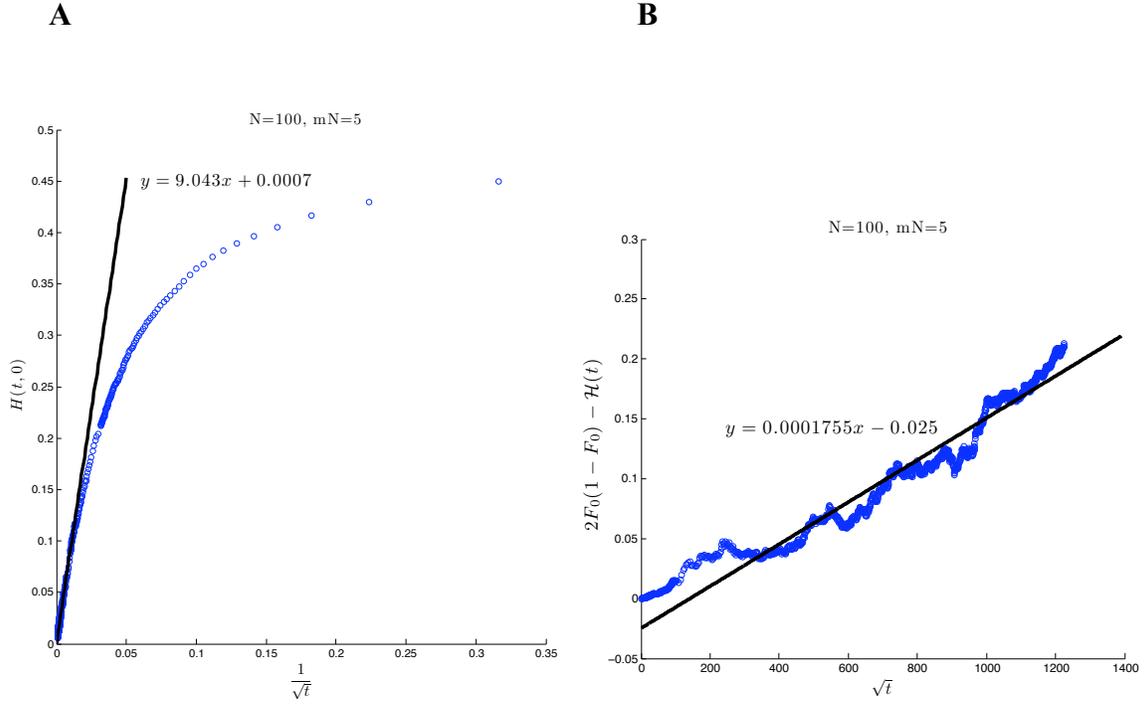

FIGURE S6 Parameter fitting of ($N$ = 100, $mN$ = 5) simulation set. (*A*) The slope of the best fit line for $t^{-1/2} \leq 0.007$ was used to estimate $D_s/D_g^2$ according to Eq. S13. (*B*) The slope of the line for intermediate $t^{1/2}$ was used to calculate $D_s$ according to Eq. S11.

**Calculation of selective coefficient *s***

Because selection acts over different $T_{exp}$ for each of the simulation sets, yet all sets reflect the same set of experiments, the selective coefficient *s* will be different for each simulation set. Just as for the diffusion constants, we can connect the slope of sector boundaries in experiments and simulations through a dimensionless invariant quantity, and the simulation boundary slopes determine *s* for each simulation set.

The experimental quantity $\frac{v_\perp}{v_\parallel} = 0.054$ (Fig. 5*C*) predicts the outward velocity of sector boundaries in simulations. Thus, the average experimental sector boundary of a sector boundary emerging from the origin would have slope $\frac{v_\parallel}{v_\perp} \approx 18.40$. To account for the fact that the experiments have an inherent $v_\parallel$ while the simulations do not, we divide the slope by $v_\parallel$ to get $v_\perp^{-1}$:

$$v_\perp^{-1} = \frac{18.40}{4.6 \times 10^{-3} \, \mu m/sec} = 4000 \frac{sec}{\mu m}. \tag{S14}$$

We multiply $v_\perp^{-1}$, which has units of sec/μm, by distance/time to create the dimensionless invariant quantity. The experimental distance is the average colony circumference ($L_{exp}$ =16,000 μm) and time is 4 days ($T_{exp}$ =345,600 sec). The relevant dimensionless invariant is therefore:

$$Inv_{exp} = \frac{v_\perp^{-1} L_{exp}}{T_{exp}} \approx 185. \tag{S15}$$

With $T_{sim}$ and $L_{sim}$ from Table S4, we can compute $v_\perp$ for each of the simulation sets from the experimental invariant quantity:

$$Inv_{exp} = Inv_{sim} = \frac{v_\perp L_{sim}}{T_{sim}} \Rightarrow v_\perp = \frac{Inv_{exp} T_{sim}}{L_{sim}} = \frac{185 \, T_{sim}}{L_{sim}}. \tag{S16}$$

| | $v_\perp$ of sector boundaries | | |
|---|---|---|---|
| **Simulation set** | $T_{sim}$ | $L_{sim}$ | $v_\perp$ |
| $N = 30$, $mN = 1$ | 252 | 200 | 233 |
| $N = 30$, $mN = 10$ | 938 | 900 | 193 |
| $N = 100$, $mN = 5$ | 4295 | 1015 | 783 |

TABLE S3 **Parameter estimation of selective coefficient** $v_\perp$ of sector boundaries from Eq. S16 for each simulation set, which are used to determine the selective coefficient *s*.

To match these $v_\perp$ values, a sector of 200 demes of F$^+$ cells was flanked with demes of F$^-$ cells. For a range of s > 0 values, the number of generations until the F$^+$ sector was annihilated by the surrounding F$^-$ sectors was averaged over 200 realizations. The simulations had a conjugation rate of $r = 0$ but underwent birth/death and migration as usual. The $v_\perp$ of the sector boundary was calculated as:

$$v_\perp = \frac{Generations\ until\ annihilation}{100} \tag{S17}$$

since each sector boundary traversed an average of 100 demes before annihilation. The $s$ values that yielded the closest observed value of Eq. S17 to the expected $v_\perp$ from Table S3 are reported in Table S4.

**Calculation of conjugation rate *r***

With $T_{sim}$, $L_{sim}$, and the selective coefficient having been established, we determined the conjugation rate *r* for each simulation set that matched the experimental circumference proportion of transconjugants with (1.6%) and without (5.2%) g3p. Simulations for each set were performed for a range of candidate *r* values; 2000 realizations were run and averaged at each potential *r* value for the $N = 30$, $mN = 1$ set, 400 realizations for $N = 30$, $mN = 10$, and 40 realizations for $N = 100$, $mN = 5$. The *r* values that most closely matched the desired circumference proportions are reported in Tables S4 and S5.

We found that the model's conjugation rate, which describes the probability of conjugation upon donor-recipient mating pair formation, can always be adjusted to accurately predict the macroscopically observed size of transconjugant domains. For example, for $N = 100$, $mN = 5$, this approach resulted in the following estimates of *r*: $r_{no\ g3p} = 1.7 \times 10^{-4}$ and $r_{g3p} = 5.5 \times$ $10^{-5}$, which is a 68% decrease in the simulation conjugation rate due to conjugation inhibition by g3p protein (Main Text Table 2). This reduction is comparable to the change in the circumference proportion of transconjugants, which was 5.2% without g3p and 1.6% with g3p. Even more important, we found that, the number of transconjugant sectors in simulations and experiments agreed (Tables S6 and S7), although only the transconjugant circumference proportion was used to estimate the conjugation rates used in simulations.

| Parameters for each simulation set | | | | | | |
|---|---|---|---|---|---|---|
| Simulation set | $D_g$ | $D_s$ | $T_{sim}$ | $L_{sim}$ | $s$ | $r$ |
| $N=30, mN=1$ | 0.066 | 0.033 | 252 | 200 | $4.8 \times 10^{-3}$ | $2.63 \times 10^{-3}$ |
| $N=30, mN=10$ | 0.080 | 0.18 | 938 | 900 | $1.5 \times 10^{-3}$ | $5.6 \times 10^{-4}$ |
| $N=100, mN=5$ | 0.019 | 0.048 | 4295 | 1015 | $4.5 \times 10^{-4}$ | $1.7 \times 10^{-4}$ |

TABLE S4 **Parameters $D_g$, $D_s$, $T_{sim}$, $L_{sim}$, $s$ and $r$ for each simulation set** The diffusion constants were calculated from the fits in Figs. S4-6. The $T_{sim}$ and $L_{sim}$ that correspond to 4 days of experimental growth for a 1μL inoculum for each simulation set were calculated according to Eqs. S8 and S9. The selective coefficient $s$ was determined as previously described, and the conjugation rate $r$ was fit to match the experimental transconjugant circumference proportion in the absence of g3p inhibition. See Table S5 for a comparison of simulation conjugation rates with and without inhibition by g3p.

| Simulation set | | $r \times T_{sim}$ | $r$ / generation |
|---|---|---|---|
| $N = 30$, $mN = 1$ | No g3p | 0.66 | $3.4 \times 10^{-3}$ |
| | g3p | 0.21 | $1.0 \times 10^{-3}$ |
| $N = 30$, $mN = 10$ | No g3p | 0.53 | $2.7 \times 10^{-3}$ |
| | g3p | 0.12 | $6.0 \times 10^{-4}$ |
| $N = 100$, $mN = 5$ | No g3p | 0.73 | $3.8 \times 10^{-3}$ |
| | g3p | 0.24 | $1.2 \times 10^{-3}$ |

**Conjugation rate for each simulation set**

TABLE S5 **Conjugation rate $r$ for each simulation set, with and without inhibition by g3p** Conjugation rates $r$ from simulations were inferred by matching the proportion of transconjugants at $T_{sim}$ to the experimental values with and without g3p. The inferred $r$/generation assumes a 30 minute generation time over the 4 day growth period.

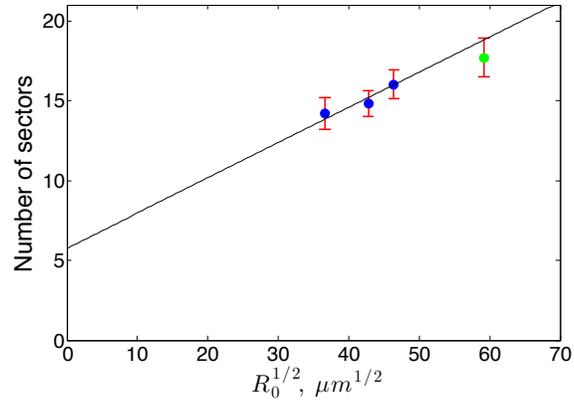

FIGURE S7 Number of surviving sectors as a function of initial radius. Same as Fig. 5*B*, but with a different set of 1, 2, 3, and 4 µL colonies imaged after 14 days. The line of best fit $y = 0.22x + 5.785$ was calculated from the 1, 2, and 3 µL drop sizes (blue) constrained to a slope of 0.22/µm$^{1/2}$ as determined from previous calculation of $\frac{D_s}{v_\parallel}$ (Fig. 5*A*).

**A** $N = 30$, $mN = 1$

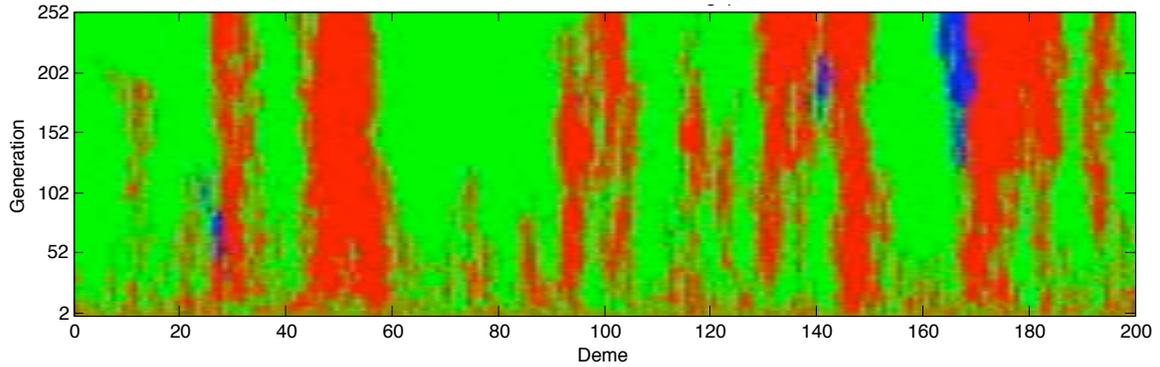

**B** $N = 30$, $mN = 10$

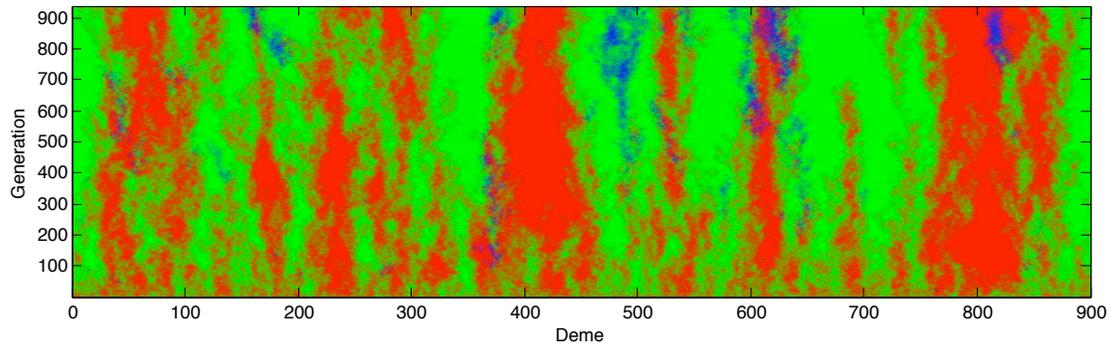

**C** $N = 100$, $mN = 5$

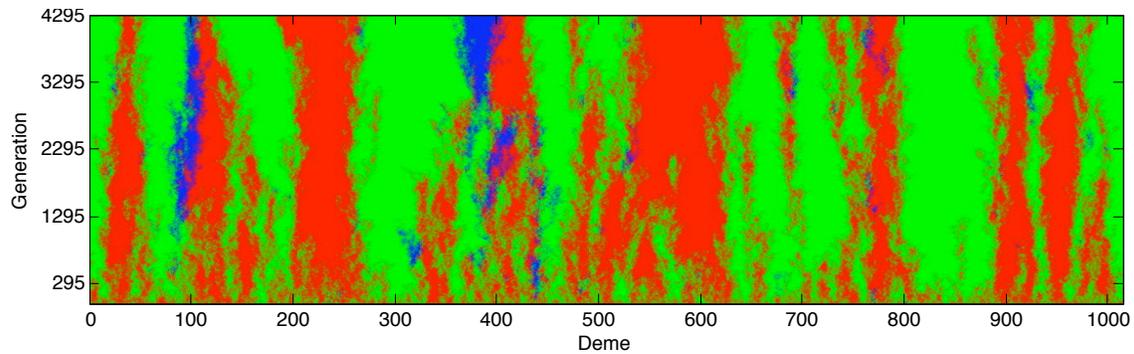

FIGURE S8 Representative visualizations of each simulation set. Parameters correspond to those calculated in the absence of g3p (Table S4). $F^+_c$ cells in red, $F^-$ in green, and transconjugants in blue.

(*A*) $N = 30$, $mN = 1$: $T_{sim} = 252$ generations, $L_{sim} = 200$ demes, $s = 4.8 \times 10^{-3}$, and $r = 2.63 \times 10^{-3}$.

(*B*) $N = 30$, $mN = 10$: $T_{sim} = 938$ generations, $L_{sim} = 900$ demes, $s = 1.5 \times 10^{-3}$, and $r = 5.6 \times 10^{-4}$.

(*C*) $N = 100$, $mN = 5$: $T_{sim} = 4295$ generations, $L_{sim} = 1015$ demes, $s = 4.5 \times 10^{-4}$, and $r = 1.7 \times 10^{-4}$. Same as Main Text Fig. 4C.

| Number of sectors in experiments and simulations | | | |
|---|---|---|---|
| Simulation set/Experimental | Transconjugant | $F^+$ | $F^-$ |
| Experimental | 2.14 ± 0.31 | 8.02 ± 0.35 | 7.86 ± 0.43 |
| $N = 30$, $mN = 1$ | 2.05 ± 0.30 | 7.95 ± 0.35 | 8.00 ± 0.32 |
| $N = 30$, $mN = 10$ | 1.80 ± 0.19 | 6.90 ± 0.20 | 7.05 ± 0.20 |
| $N = 100$, $mN = 5$ | 2.00 ± 0.27 | 8.00 ± 0.27 | 8.00 ± 0.24 |

TABLE S6 **Number of sectors in experiments and simulations without inhibition by g3p** The mean number of sectors ± SE for experiments reported in Fig. S3. For each simulation set, 20 realizations were run with the respective $r$ without g3p from Table S5. The number of each type of sector was visually inspected and averaged over the realizations ± SE.

| Simulation set/Experimental | Transconjugant | $F^+$ | $F^-$ |
|---|---|---|---|
| Experimental | $1.00 \pm 0.22$ | $11.14 \pm 0.47$ | $10.86 \pm 0.40$ |
| $N = 30$, $mN = 1$ | $1.00 \pm 0.13$ | $8.85 \pm 0.38$ | $9.00 \pm 0.41$ |
| $N = 30$, $mN = 10$ | $0.30 \pm 0.10$ | $6.85 \pm 0.53$ | $7.00 \pm 0.48$ |
| $N = 100$, $mN = 5$ | $1.00 \pm 0.26$ | $9.00 \pm 0.28$ | $9.00 \pm 0.26$ |

**Number of sectors in experiments and simulations with inhibition by g3p**

TABLE S7 **Number of sectors in experiments and simulations with inhibition by g3p** The mean number of sectors ± SE for experiments plotted in Fig. S3. For each simulation set, 20 realizations were run with the respective *r* with g3p from Table S5. The number of each type of sector was visually inspected and averaged over the realizations ± SE.

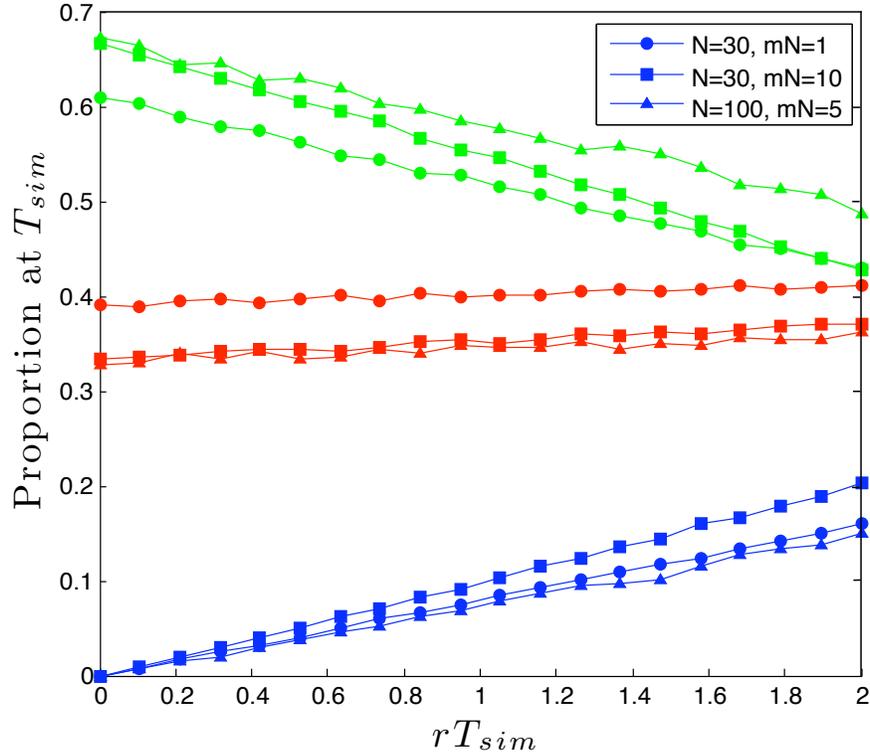

FIGURE S9 Model-based quantitative predictions of population dynamics over a range of conjugation rates. Here each of the three simulation sets was run for its respective $T_{sim}$ and $s$ in Table S4. So that the different sets are comparable, the conjugation rates $r$ on the x-axis are scaled by $T_{sim}$. Because the starting populations were originally 50% $F^+_c$ and 50% $F^-$, the frequencies at $rT_{sim} = 0$ are a result of selection alone (no conjugation). The $rT_{sim}$ values representing 4 days of growth without g3p (transconjugant proportion 5.2%) for the three simulation sets are 0.66; 0.53; and 0.73, respectively, so the largest $rT_{sim}$ corresponds to a conjugation rate approximately three times larger than observed in our experiments. The ($N = 30$, $mN = 1$) and ($N = 30$, $mN = 10$) simulation sets were each averaged over 2000 realizations while the ($N = 100$, $mN = 5$) set was averaged over 300 realizations. $F^+_c$ cells labeled as red, $F^-$ as green, and transconjugants as blue.

# SUPPORTING REFERENCES